\documentstyle[prd,aps]{revtex} 

\input psfig \pssilent
\begin{document}
\title{Discrete quantum gravity: a mechanism for selecting
the value of fundamental constants
\footnote{This essay received an "honorable mention"
        in the 2003 Essay Competition of the Gravity
        Research Foundation -- Ed.}}

\author{\bf Rodolfo Gambini$^1$, and Jorge Pullin$^2$}
\address{1. Instituto de F\'{\i}sica, Facultad de Ciencias, 
Universidad
de la Rep\'ublica\\ Igu\'a esq. Mataojo, CP 11400 Montevideo, Uruguay}
\address{2. Department of Physics and Astronomy, 
Louisiana State University,\\ 202 Nicholson Hall, Baton Rouge,
LA 70803-4001}

\date{June 20th 2003}
\maketitle

\begin{abstract}
Smolin has put forward the proposal that the universe
fine tunes the values of its physical constants through a Darwinian
selection process. Every time a black hole forms, a new universe is
developed inside it that has different values for its physical
constants from the ones in its progenitor. The most likely universe is
the one which maximizes the number of black holes. Here we present a
concrete quantum gravity calculation based on a recently proposed
consistent discretization of the Einstein equations that
shows that fundamental physical constants change in a random fashion
when tunneling through a singularity. 
\end{abstract}

Fundamental constants in nature need to fall within a rather narrow
set of values for the universe to have its current form, in particular
to accommodate life. When written in dimensionless form, unnaturally
large ratios appear between various of the fundamental
constants. Inflation has been proposed as a mechanism to account for
several features of the universe, but it cannot explain the values of
all fundamental physical constants nor provide a complete resolution
to the hierarchy problem. A recent proposal due to Smolin \cite{Sm}
poses that the selection of the values of the fundamental physical
constants happens through a Darwinian process. Whenever a universe
develops a black hole a new universe forms within it with different
values of the physical constants. The universes that are naturally
selected are those such that the physical constants are such that they
maximize the likelihood of formation of black holes. That allows such
universes to reproduce more efficiently. This attractive proposal has
the feature that it can be tested experimentally. It can be falsified
by showing that the values of the physical constants are such that we
are not at the maximum likelihood of formation of black holes. The
proposal has been recently reconsidered by Bjorken \cite{Bj}. Several
criticisms have been levied against these arguments. One of the
problems up to now has been the lack of a detailed mechanism to
account for the change of fundamental physical constants during the
tunneling through a black hole.

In this paper we would like to discuss a detailed scenario in which
changes in the fundamental physical constants can occur when tunneling
through a singularity. Having a detailed scenario for tunneling might
be interesting cosmologically even independently from Smolin's Darwinian
hypothesis \cite{Br}.

The proposed scenario is based on the recently introduced consistent
discretization technique for treating quantum general relativity on
the lattice \cite{GaPu}. The technique constructs a discrete theory on
the lattice that represents an approximation to general relativity and
such that all of its equations can be solved simultaneously (usual
discretizations of general relativity produce an inconsistent set of
equations). The discrete theories constructed with the new technique
have several attractive features.  Among them is the presence of well
understood symmetries that provide a lattice representation of the
symmetries of general relativity \cite{cosmo}. This is quite novel,
since it has been a long standing problem how to reconcile the
continuous coordinate invariance of general relativity with the 
discreteness  of a lattice framework.

In this letter we analyze a concrete example of a bounce through a
singularity in the consistent lattice approach. We will consider a
Friedman universe with a cosmological constant and a (very massive)
scalar field. This is the simplest model we have found that exhibits
bounce through a singularity. As discussed in \cite{cosmo},
anisotropic models exhibit similar behavior. The approach to the
singularity in the interior of a black hole can be modeled as an
anisotropic cosmology and therefore the following discussion is of
relevance to the behavior of the interior of a black hole. It should
be noticed that several mechanisms have been postulated in the past
for tunneling through a black hole. Some of these have been classical,
postulating the development of a cosmological constant in the interior
\cite{cl}, quantum inspired modifications of general relativity
\cite{Br}, or path integral formulations of quantum gravity
\cite{HaHa}, but most of these have not been associated with changes
in the fundamental physical constants, although see \cite{ahdy}.

The Lagrangian for the model, written in terms of Ashtekar's variables
\cite{kodama} is,
\begin{equation}
L= E\dot{A} + \pi \dot{\phi}- N E^2 (-A^2+(\Lambda +m^2 \phi^2)|E|)\label{lag}
\end{equation}
where $\Lambda$ is the cosmological constant, $m$ is the mass of the
scalar field $\phi$, $\pi$ is its canonically conjugate momentum and
$N$ is the lapse with density weight minus one. The appearance of
$|E|$ in the Lagrangian is due to the fact that the term cubic in $E$
is supposed to represent the spatial volume and therefore should be
positive definite. In terms of the ordinary lapse $\alpha$ we have
$\alpha = N |E|^{3/2}$. 

We consider the evolution parameter to be a discrete variable. Then the
Lagrangian becomes
\begin{equation}
L(n,n+1)=E_n (A_{n+1}-A_n)+ \pi_n (\phi_{n+1}-\phi_n) -N_n E_n^2 (-A^2_n+
(\Lambda+m^2 \phi^2_n) |E_n|)
\end{equation}
The discrete time evolution is generated by a canonical transformation of
type 1 whose generating function is given by $-L$, viewed as a function
of the configuration variables at instants $n$ and $n+1$. Evolution 
equations can be written for all the variables and their canonical
momenta. The evolution equations are made consistent by determining
the Lagrange multipliers $N_n$. The resulting equations can be 
reduced to \cite{cosmo},
\begin{eqnarray}
P^A_{n+1}&=&A_n^2 \Theta^{-1}\label{ecupa}\\
A_{n+1}&=&{3A_n^2-P^A_n\Theta\over 2A_n}\label{ecua}\\
\phi_{n+1}&=&\phi_n\\
P^\phi_{n+1}&=&P^\phi_n -\left(A_n^3-P^A_n\Theta A_n\right)m^2\phi_n\Theta^{-2}
\end{eqnarray}
where $\Theta=\Lambda+m^2\phi_n^2$. It should be noted that these
equations preserve the symplectic structure, that is, the variables
$(P^A_{n+1},A_{n+1})$ have the same canonical Poisson brackets as
$(P^A_n,A_n)$. To make contact with the variables of the continuum, we
note that the triad $E_n=P^A_{n+1}$.

\begin{figure}
\centerline{\psfig{file=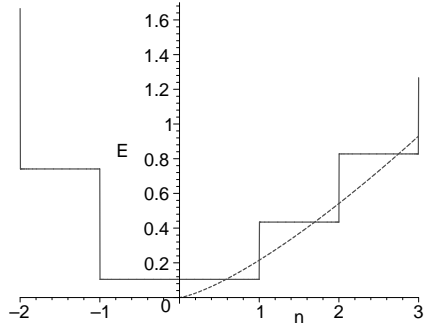,height=5cm}}
\caption{The approach to the singularity in the discrete and continuum 
solutions. The discrete theory has a small but non-vanishing triad at
$n=0$ and the singularity is therefore avoided.}
\label{fig1}
\end{figure}

The discrete evolution equations have the feature that they avoid
the singularity present in the continuum model for generic sets of
initial data \cite{sing}. In figure \ref{fig1} we show a generic
evolution near the region where classically one would encounter
a singularity. One can see that the discrete theory, although 
approximating reasonably well the continuum behavior does not have
the metric going through zero.

It should be noted that the resulting theory has no constraints,
unlike the continuum theory \cite{Ahlu}. Therefore one does not
confront the problem of finding ``observables'', that is, quantities
that have vanishing Poisson brackets with the constraints. The
discrete theory has four phase space degrees of freedom and one can
introduce four constants of motion. Three of these constants of motion
do not depend explicitly on the evolution parameter $n$.  Remarkably,
two of them can be viewed as discretizations of the two independent
observables of the continuum theory. Therefore the discrete theory has
in a precise sense embedded in it the symmetries of the continuum
theory \cite{cosmo}. The symmetries are present in the discrete theory
in the sense that there exist constants of the motion independent of
the evolution parameter that one can use to generate canonical
transformations representing the symmetries.

The remaining two constants of motion vanish in the continuum limit.
Let us concentrate on the one that is independent of the evolution
parameter. It arises from considering the canonical transformation
that generates time evolution as an exponentiation of a quantity that
plays a role of generalized Hamiltonian for the discrete model (the
model does not have a genuine Hamiltonian since time is discrete). The
generalized Hamiltonian should therefore be preserved under
evolution. If one recasts the evolution equation as,
\begin{equation}
A_{n+1}=A_n+ \{A_n,H_n\}+ {1\over 2!} \{\{A_n,H_n\},H_n\}+
\ldots
\end{equation}
and similarly for the other variables, 
one can read off the ``Hamiltonian'' 
\begin{equation}
H_n = {{\cal C}_n^2\over 4 \Theta A_n}\left[1+\sum_{k=1}^\infty a_k 
\left({{\cal C}_n \over A_n^2}\right)^k\right]\label{ham}
\end{equation}
where ${\cal C}_n=\left(A_n^2-P^A_n\Theta \right)$ is the
discretization of the Hamiltonian constraint of the continuum theory
and $a_1=1/(3\times 4),a_2=1/(6\times4^2),a_3=0,a_4=-1/(6\times 4^4),
a_5=-1/(15\times 4^5),a_6=7/(10\times 4^6)$ etc. The power series
nature of the definition of this constant implies that it exists only
where the series is convergent. Whenever it is, the finite canonical
transformation can be written as an exponentiation of an infinitesimal
canonical transformation (contact transformation). To understand this,
notice that the canonical transformation that materializes the
discrete evolution is singular when $A_n=0$ (see equation
(\ref{ecua}).  This singularity separates the phase space into two
disjoint regions $A_n>0$ and $A_n<0$. The contact transformation will
fail to exist when the canonical transformation connects points that
lie different disjoint regions, i.e. when ${\rm sg}(A_{n+1}) \neq {\rm
sg}(A_{n})$. This happens when $3A_n^2-P^A_n\Theta<0$. This in turn
implies that the expansion parameter of the series (``normalized
constraint'') $|{\cal C}_n/A_n^2|>2$.  The prediction is therefore that
this will be a conserved quantity until the evolution takes us over
the singularity in the canonical transformation.

\begin{figure}
\centerline{\psfig{file=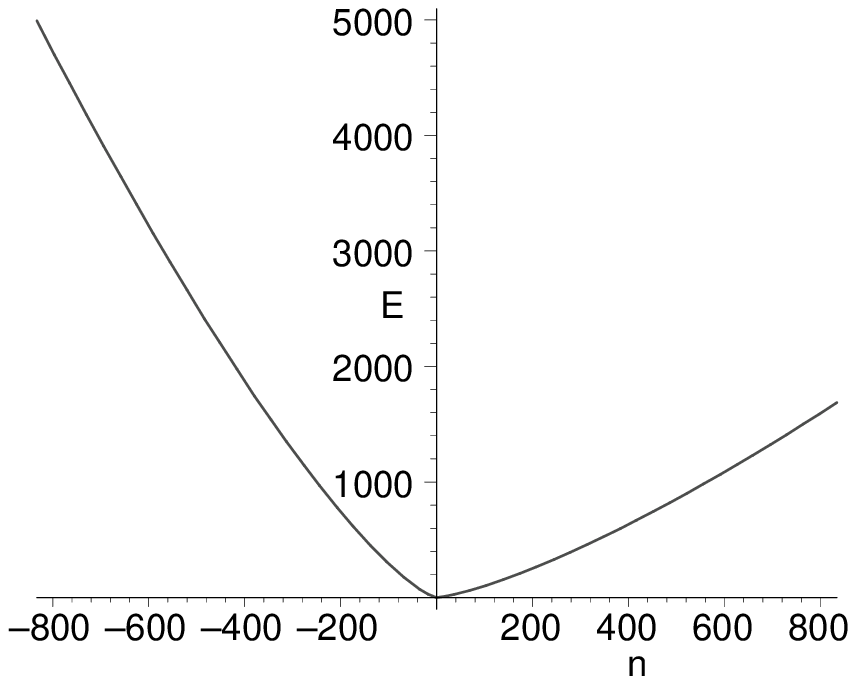,height=5cm}
\psfig{file=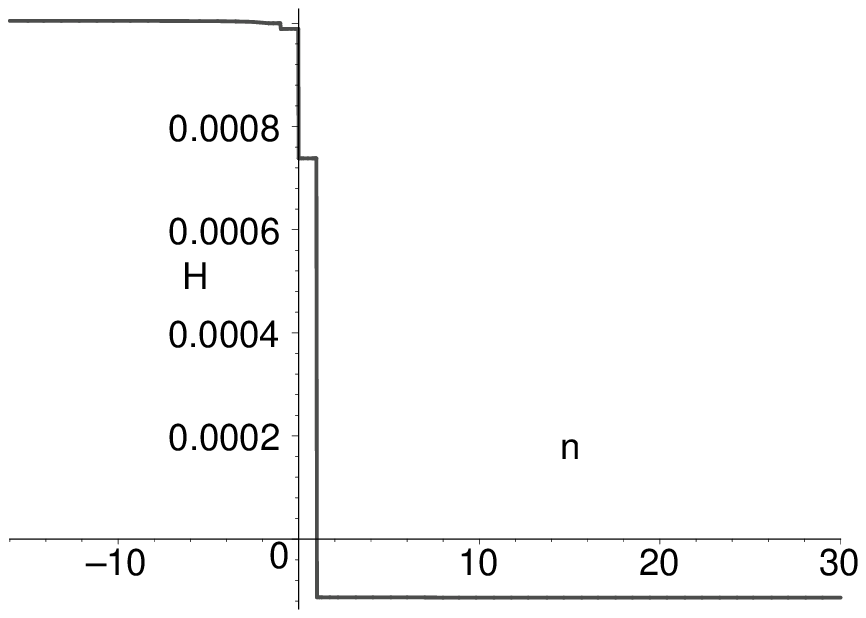,height=5cm}}
\caption{Typical behavior of the discrete evolution of the triad $E$ 
and the constant of
motion $H$. One can see that the rate of expansion/contraction differs
when tunneling through the singularity. This is reflected itself in 
a jump in the value of the constant of the motion $H$.} 
\label{fig2}
\end{figure}

\begin{figure}
\centerline{\psfig{file=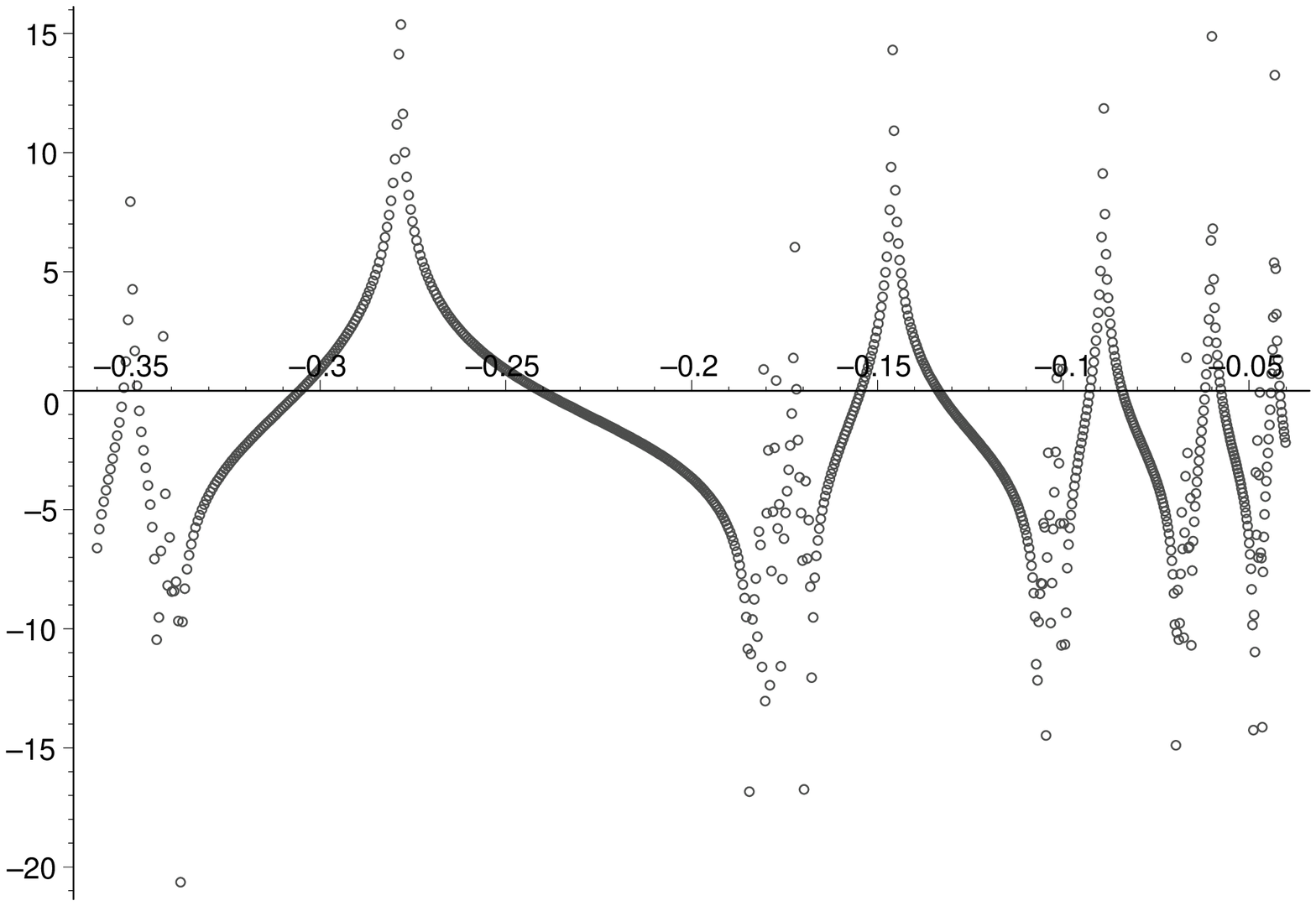,height=7cm}}
\caption{The value of the logarithm of the observable after the bounce
as a function of the initial value of the observable. We see that there exist
ranges in which very small variations of the initial value translate themselves in large 
changes in the final value. This shows that the change in the fundamental constants
in black hole tunneling is not deterministic, leading to a Darwinian picture.}
\label{fig3}
\end{figure}

As seen in figure \ref{fig2} the rate of expansion/contraction changes
when tunneling through the singularity. The constant of the motion can
be seen as an invariant characterization of such a rate. More
precisely, the constant of the motion, which vanishes in the continuum
limit, is a measure of how well the discrete theory is approximating
the continuum one. What changes in the tunneling is the lattice 
spacing.
It is worthwhile noticing that the magnitude of the jump in the
constant of the motion exhibits sensitive dependence on the initial
condition of the problem. A small change in the initial values can
lead to large changes in the behavior of the jump.  To view it in
another words, the rate of expansion after the tunneling loses
correlation with respect to the rate of contraction before the
tunneling. This can be seen in figure \ref{fig3}.

At this point the reader may wonder what is the connection between the
calculation we presented of the ``bounce'' in the value of an
observable in a model cosmology and the values of the fundamental
physical constants. Of course, a detailed model involving interacting
fields and local degrees of freedom will require a much larger
calculational effort than what we are able to attempt at present. At
this point, we can only present a heuristic argument.  The argument is
based on the fact that the observable we discussed is an invariant
measure of the lattice separation. Therefore its value is connected
with how refined the lattice spacing in the theory is. In this
particular model, the lattice is only in the time-like direction, but
one can expect that in more realistic models, with local degrees of
freedom, similar behaviors will occur for the spatial lattice
spacings. Now, in a lattice gauge theory the values of the fundamental
physical constants are related to the bare values that appear in the
Lagrangian through a limiting process in which one takes the lattice
spacing to zero and also fine-tunes the bare parameters in such a way
that the ``dressed'' physical constants are finite. More precisely,
such a process is fine tuned for one observable, and then the same
process predicts the values of other observables (at least for
renormalizable theories). In the gravitational case we do not expect
to have renormalizability in the traditional sense, so the proposal we
are presenting is that the theory remains discrete. In the discrete
theory there will exist states that approximate the continuum theory
better than others and a measure of this will be given by the value of
the observable we discuss. The value of the ``dressed'' physical
constants will depend therefore on the bare values and on the value of
the ``spacing''. An invariant measure of the ``spacing'' in the
lattice is given by the value of the ``Hamiltonian'' (\ref{ham}). In a
situation with ``fine'' spacing its value will be small.  In this
scenario, the tunneling through the singularity we have exhibited will
translate itself in a change in the value of the ``Hamiltonian''
and therefore in a change in the value of the fundamental constants.

The discussion up to now has been entirely classical. However, the
discontinuity in the constant of the motion has a quantum counterpart.
To quantize the system, one represents the quantum evolution via a
unitary transformation that implements at the level of Heisenberg
equations the discrete equations of motion associated with the
canonical transformation we discussed above. We will not give the
details here, some of them can be seen in \cite{cosmo}. In the
quantum theory, one can consider a state peaked around a classical
solution and evolve it using the discrete evolution operator. One
cannot directly promote the ``Hamiltonian'' to a quantum self-adjoint
operator since it is not well defined near the bounce. However, one
can take a finite number of terms of its expansion. This will
approximate well (classically) the behavior of the ``Hamiltonian'' far
away from the bounce. Such a finite expansion can be promoted to a
self-adjoint quantum operator. This allows, for instance, to compute
its expectation value before and after the point where one would have
expected the big bang classically. Generically the values will be
different, mirroring what we found in the classical theory.

The singularity that arises in the big bang has elements in common
with the singularity in the interior of black holes. Although the
cosmological model we studied in detail is not  the one that has
elements in common with the black hole interior, the feature we
presented of tunneling through the singularity is expected to exist in
a variety of models. Also, the details of how the changes occur could
differ with different discretization schemes. We can therefore expect
a similar phenomenon to be present in the interior of black holes,
although the details may vary. Each black hole will have its
singularity replaced by tunneling into a new universe, in which the
dressed value of the fundamental constants will be different.  The
change in the values has elements of randomness in it. This allows to
construct a picture of the universe in which ``evolution'' takes place
every time a black hole is formed, as was the original proposal of
``The life of the cosmos'' \cite{Sm}.

We wish to thank D. Ahluwalia for comments.
This work was supported by grants NSF-PHY0090091,
funds of the Horace Hearne Jr.  Institute for Theoretical Physics, the
Fulbright Commission in Montevideo and PEDECIBA (Uruguay).


\begin{references}
\bibitem{sing} In this simple model the singularity is only a 
coordinate singularity in the continuum theory, but one can consider
more elaborate models where one deals in the same fashion with a real
singularity.
\end{references}
\end{document}